\documentclass[twocolumn,showpacs,fleqn,nobibnotes]{revtex4}

\usepackage{amsmath}
\usepackage{graphicx}
\usepackage{float}
\usepackage{subfigure}

\def\lsim{\raise0.3ex\hbox{$<$\kern-0.75em\raise-1.1ex\hbox{$\sim$}}}
\def\gsim{\raise0.3ex\hbox{$>$\kern-0.75em\raise-1.1ex\hbox{$\sim$}}}

\begin{document}
\newcommand\ie {{\it i.e.}}
\newcommand\eg {{\it e.g.}}
\newcommand\etc{{\it etc.}}
\newcommand\cf {{\it cf.}}
\newcommand\etal {{\it et al.}}
\newcommand{\be}{\begin{eqnarray}}
\newcommand{\ee}{\end{eqnarray}}
\newcommand{\jp}{$ J/ \psi $}
\newcommand{\pp}{$ \psi^{ \prime} $}
\newcommand{\ppp}{$ \psi^{ \prime \prime } $}
\newcommand{\dd}[2]{$ #1 \overline #2 $}
\newcommand\noi {\noindent}

\title{Nuclear shadowing and prompt photons at relativistic hadron colliders}
\pacs{13.85.Qk; 12.38.-t}
\author{C. Brenner Mariotto $^{a}$ and V.P. Gon\c{c}alves
$^{b}$}

\affiliation{
$^a$ Departamento de F\'{\i}sica, Universidade Federal do Rio Grande \\
Caixa Postal 474, CEP 96201-900, Rio Grande, RS, Brazil
\\
$^b$ Instituto de F\'{\i}sica e Matem\'atica, Universidade Federal de
Pelotas
\\
Caixa Postal 354, CEP 96010-900, Pelotas, RS, Brazil\\
}

\begin{abstract}
The production of prompt photons at high energies  provides a direct probe of the dynamics of the strong interactions. In particular, one expect that it could be used to constrain the  behavior of the nuclear gluon distribution in $pA$ and $AA$ collisions. In this letter we investigate the influence of nuclear effects in the production of prompt photons and estimate the transverse momentum dependence of the nuclear ratios $R_{pA}  = { \frac{d\sigma (pA)}{dy d^2 p_T} } / A {\frac{d\sigma
(pp)}{dy  d^2 p_T}}$ and $R_{AA}  = { \frac{d\sigma (AA)}{dy d^2 p_T} } / A^2 {\frac{d\sigma
(pp)}{dy d^2 p_T}}$ at RHIC and LHC energies.  We demonstrate that the study of these observables can be useful to determine the magnitude of the shadowing and antishadowing effects in the nuclear gluon distribution.

\end{abstract}

\maketitle
In the last years the study of heavy ion collisions have
provided strong evidence for the formation of a quark-gluon plasma (QGP) \cite{sps,rhic}, which
is one of the main predictions of the Quantum Chromodynamics (QCD).  Currently, distinct models associated to different assumptions describe reasonably the experimental data, with the main uncertainty present in these analysis directly connected with the poor knowledge of the initial
conditions of the heavy ion collisions. Theoretically, the early
evolution of these nuclear collisions is governed by the dominant
role of gluons, due to their large interaction
probability and the large gluonic component in the initial nuclear
wave functions.  Such extreme condition is
expected to significantly influence QGP signals and should modify
the hard probes produced at early times of the heavy ion
collision. Consequently, a systematic measurement of the nuclear
gluon distribution is of fundamental interest in understanding the parton structure of nuclei and to determine the initial
conditions of the QGP. Other important motivation for the
determination of the nuclear gluon distributions is that the high
density effects expected to occur in the high energy limit of QCD
should be manifest in the modification of the gluon dynamics.

\begin{figure}[t]
\includegraphics[scale=0.35]{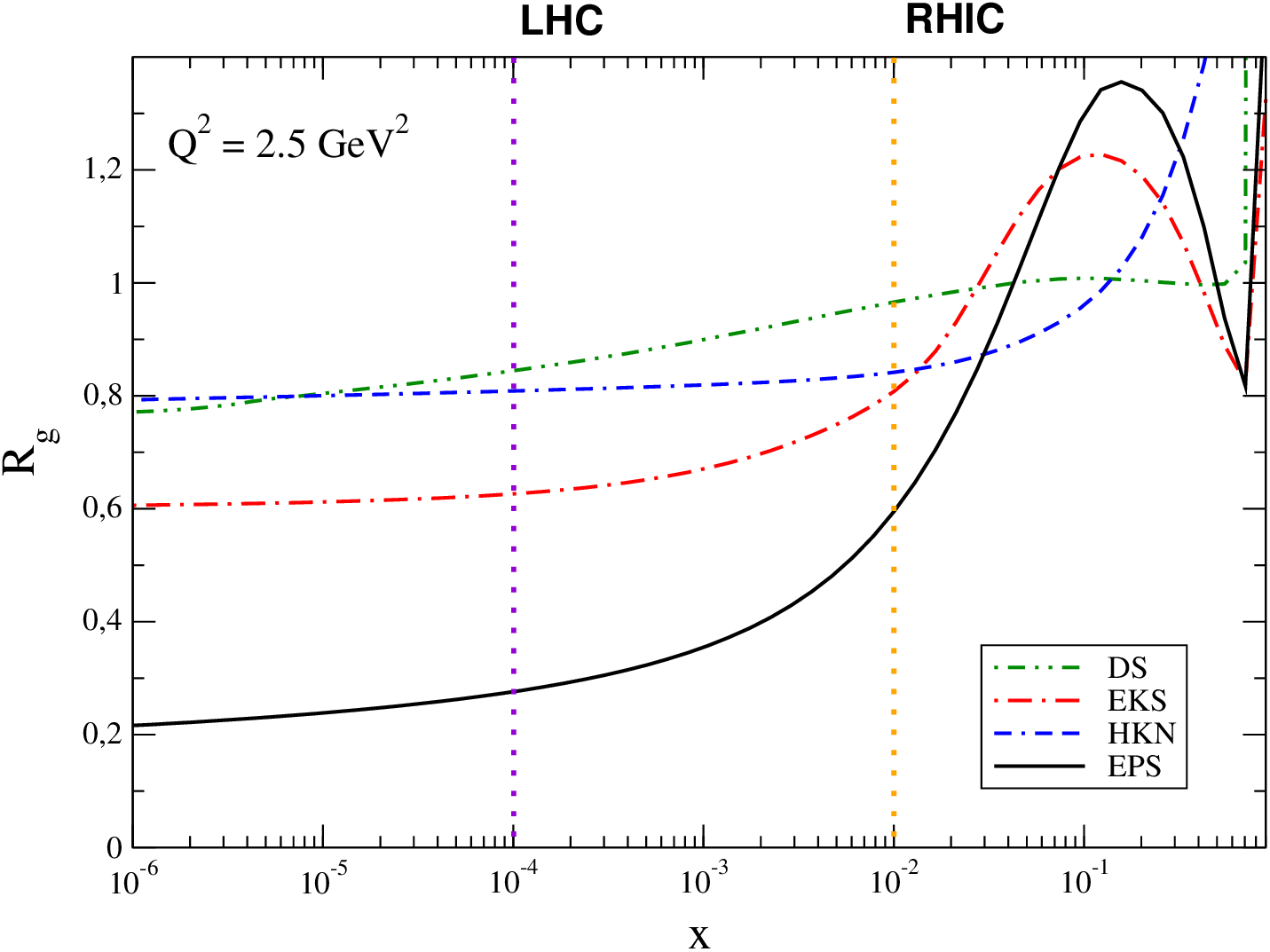}
\caption{Ratio $R_g \equiv xg_A/A.xg_N$ predicted by the  DS \cite{sassot}, EKS \cite{EKS98}, HKN \cite{HKN} and EPS \cite{EPS08} parameterizations at $Q^2 = 2.5$ GeV$^2$ and $A = 208$. The vertical lines indicate the minimum $x$ values which are probed in photon production at central rapidities in proton-nucleus collisions at RHIC and LHC.} 
\label{fig1}
\end{figure}

One of the nuclear effects which is expected to modify the behavior of the nuclear gluon distribution is the nuclear shadowing (For a recent review see Ref. \cite{armesto}). This effect has been observed in the nuclear structure functions by different experimental collaborations  \cite{arneodo,e665} in the study of the deep inelastic lepton scattering (DIS) off nuclei. The modifications on $F_2^A (x,Q^2)$ 
 depend on the parton momentum fraction $x$. While for momentum fractions $x < 0.1$ (shadowing region) 
and $0.3 < x < 0.7$ (EMC region), a
depletion is observed in the nuclear structure functions, in the intermediate region ($0.1 < x < 0.3$) it is verified an enhancement known as antishadowing. These experimental results strongly constrain the behavior of the nuclear quark distributions. However, the  nuclear gluon distribution is usually inferred  using the momentum sum
rule as constraint and/or studying the log $Q^2$ slope of the ratio $F_2^{Sn}/F_2^C$ \cite{gousset}. Due to the scarce experimental data in the small-$x$ region and/or for observables strongly dependent on the nuclear gluon distribution, the current status is that its behavior is completely undefined. It is demonstrated by the analysis of the Fig. \ref{fig1}, where we present the results for the ratio  $R_g \equiv xg_A/A.xg_N$ predicted by the  EKS \cite{EKS98},  DS \cite{sassot}, HKN \cite{HKM,HKN} and EPS \cite{EPS08} parameterizations at $Q^2 = 2.5$ GeV$^2$ and $A = 208$. These four groups realize a global analysis of the nuclear experimental data using the DGLAP evolution equations \cite{dglap}. From Fig. \ref{fig1}, we have that currently the magnitude of the shadowing and the presence or not of the antishadowing  are open questions. It has motivated several authors to propose the study of different observables in distinct processes to constrain the nuclear gluon distribution (See Ref. \cite{armesto}). For instance, in Ref. \cite{vic_bert} it was proposed to study the heavy quark  and vector meson production in ultraperipheral heavy ion collisions at LHC, while in Ref. \cite{vic_caza_gluon} the study of the longitudinal and charm structure functions in $eA$ collisions at RHIC. Another possibilities are the heavy quark and quarkonium production in central proton-nucleus and nucleus-nucleus collisions (See e.g. Refs. \cite{vogt1,vogt2,vogt3,vogt4,vic_luiz1,ABMG:2006}). Furthermore, in Refs. \cite{greiner,jamal} the authors have studied the nuclear effects in prompt photon production. This process provides a direct probe of the dynamics of strong interactions and, consequently, of the nuclear gluon distribution, since the dominant mechanism of production at high energies is the Compton scattering $q + g \rightarrow \gamma + q$. More recently, the possibility to observe the nuclear gluon shadowing using high-$p_T$ prompt photon production at RHIC and at LHC was discussed in Ref. \cite{arleo}. It was argued that the production of isolated photons turns out a promising channel for the extraction of $R_g$. In this letter we extend these previous studies considering the more recent nuclear parton parameterizations and estimating the transverse momentum dependence of the ratios 
$R_{pA}  = { \frac{d\sigma (pA)}{dyd^2 p_T} } / A {\frac{d\sigma
(pp)}{dy d^2 p_T}}$ and $R_{AA}  = { \frac{d\sigma (AA)}{dy d^2 p_T} } / A^2 {\frac{d\sigma
(pp)}{dy d^2 p_T}}$ at  RHIC and LHC energies, where the differential cross section is given in Eq. (\ref{cslo}) below. Our goal is to determine if these observables can be used to constrain the different effects expected to be present in the nuclear gluon distribution and to discriminate among the distinct models present in the literature.

Prompt photon production occurs through two types of processes: 
the direct piece, where the photon is emitted via a point like coupling
to a quark, and the fragmentation piece, in which the photon
originates from the fragmentation of a final state parton. As the
second component  can be almost completely reduced by isolation
criterion used in the experimental data analysis, we focus our study
only in the direct component, which provide a clean probe of the
hard scattering dynamics. In this case, the prompt
photon production cross section is given by \cite{owens}
\begin{eqnarray}
\frac{d\sigma_{h_1 h_2 \rightarrow \gamma X}}{dydp_T^2} = \sum_{i,j,k}
\int_{x_{1}^{min}}^1 dx_1 f_i(x_1,Q^2) f_j(x_2,Q^2) \nonumber \\
\frac{x_1x_2}{2x_1-x_Te^y}
 \frac{d \hat{\sigma}_{ij\rightarrow \gamma k}}{d\hat{t}} (Q^2,x_1,x_2) \,\,,
\label{cslo}
\end{eqnarray}
where $x_T=2p_T/\sqrt{s}$, $y$ and $p_T$ are the rapidity and
transverse momentum of the produced photon, $f_i (x,Q^2)$ are the
parton densities, $x_1$ and $x_2$ are the momentum fractions of the
partons involved in the hard process. In this case we have that
$x_2=\frac{x_1x_Te^{-y}}{2x_1-x_Te^y}$ and
$x_{1}^{min}=\frac{x_Te^{y}}{2-x_Te^{-y} }$. $Q^2$ is the hard scale 
and $\frac{d\hat{\sigma}}{d\hat{t}}$ are the partonic cross
sections, which are perturbatively calculable \cite{field}. 
The contributing LO subprocesses are $qg \rightarrow q \gamma$ (Compton), $q \bar q
\rightarrow g \gamma$ (annihilation), followed by the subdominant
diagrams $q \bar q  \rightarrow \gamma \gamma$ (pure EM), $g g
\rightarrow \gamma \gamma$ and $g g  \rightarrow g \gamma$.
The correspondent LO matrix elements and partonic cross sections can
be found in Refs. \cite{owens,field}. In what follows we estimate the differential cross sections for central rapidities. It implies that  the  cross section at small values of the photon transverse momentum in $pPb$ ($dAu$) collisions at LHC (RHIC) is determined by the behavior of the nuclear gluon distribution at $x_g \gtrsim 10^{-4} \, (10^{-2})$ (See Fig. \ref{fig1}). It is important to emphasize that smaller values of $x$ contribute at photon production in the forward rapidity region which can be measured, e.g., with the PHENIX and STAR detectors at RHIC and by the CMS experiment at LHC.

The main input in the  calculations of the prompt photon cross section are the nuclear parton distribution functions (nPDF). In the last years several  groups has proposed parameterizations for the nPDF, which are based on different assumptions and techniques to perform a global fit of different sets of 
data  (in particular, EPS includes RHIC data) using the DGLAP evolution equations \cite{EKS98,sassot,HKM,HKN,EPS08}.
In Fig. \ref{fig1} we present a comparison among the distinct parameterizations for the nuclear modification factor for the gluon, $R_g(x,Q^2)$, at $A = 208$ and $Q^2 = 2.5$ GeV$^2$. As we can see, these parameterizations predict very distinct magnitudes for the nuclear effects. For larger values of $x$, the EKS and the EPS show antishadowing, while this effect is absent for the HKN and EPS parameterizations in the $x \le 10^{-1}$ domain. The more surprising feature is however the amount of shadowing in the different parameterizations. While the shadowing is moderate for DS and HKN parameterizations 
and somewhat 
bigger for EKS one, the EPS prediction has a much stronger suppression compared with the other parameterizations. For smaller $x$ around $x\simeq {10}^{-5}$, while DS and HKN parameterizations have about $20\%$ suppression and EKS one have about $40\%$ suppression, for the EPS parameterization this effect goes to almost $80\%$ suppression in the nuclear gluon compared with the $A$ scaled gluon content in the proton! For bigger values of $x$ the behavior is distinct for all parameterizations. As $x$ grows, the DS parameterization predicts that $R_g$ grows continuously to 1, that means that the shadowing dies out when $x\to 10^{-1}$. The same happens for the HKN one in this limit, but this growth starts only at $x>10^{-2}$, with $R_g$ being flat for $10^{-5}<x< 10^{-2}$. At $x\approx 10^{-1}$, we have that behaviors predicted by the EKS and EPS parameterizations are similar, with $R_g$ exceeding 1.2. The main distinction between these parameterizations is that in the EPS parameterization  one has a much steeper growth, from a much stronger suppression at smaller $x$ to the antishadowing behavior for larger values of $x$.

\begin{figure}[t]
\includegraphics[scale=0.35]{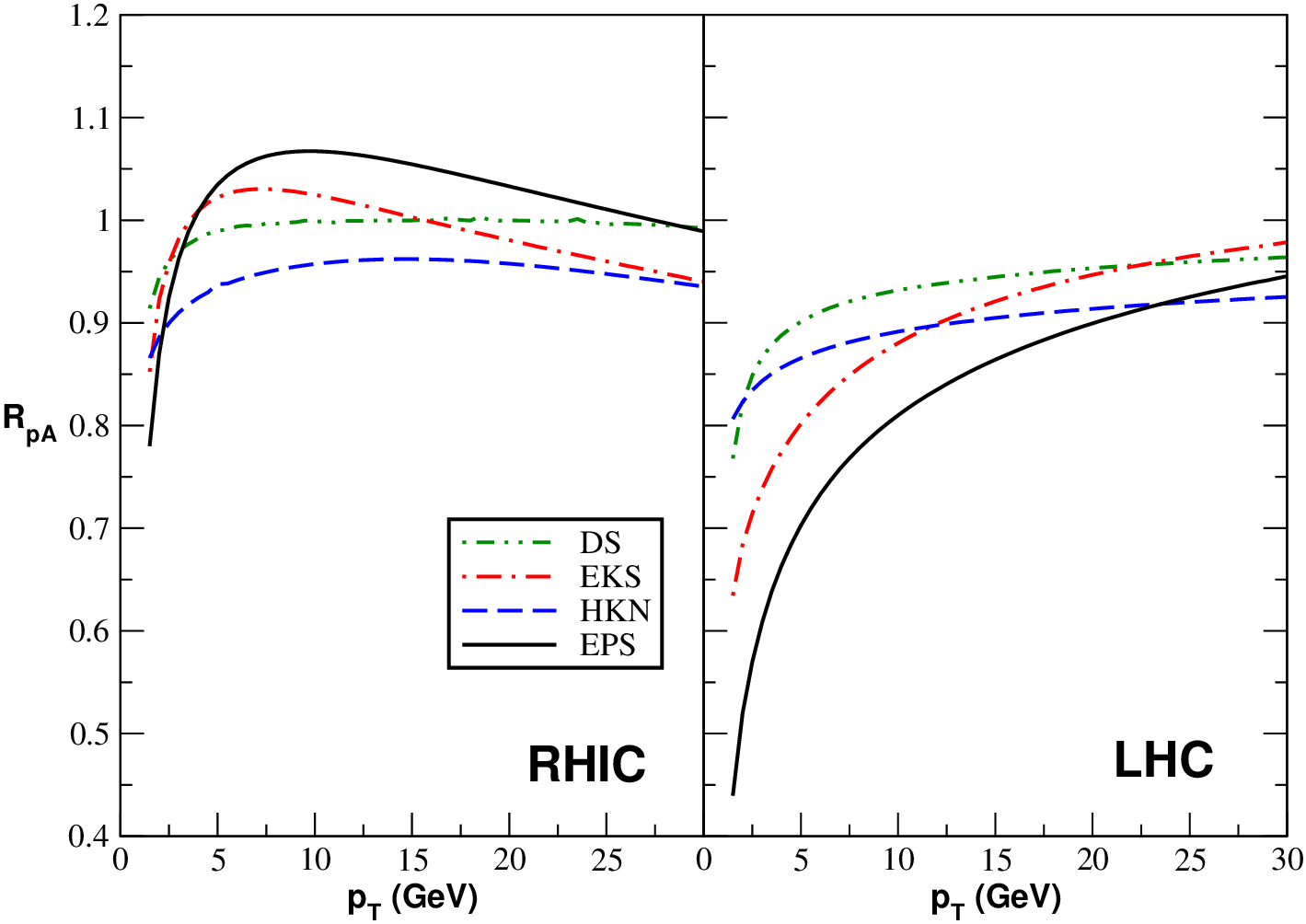}
\caption{Transverse momentum dependence of the ratio $R_{pA}$ 
in prompt photon production at RHIC ($\sqrt{s}=200\, GeV$) and LHC ($\sqrt{s}=8.8\, TeV$), for different choices of nuclear parton distributions: DS \cite{sassot}, EKS \cite{EKS98}, HKN \cite{HKN} and EPS \cite{EPS08}.
} \label{fig:RpA}
\end{figure}

In what follows we
calculate the prompt photon production cross section pA and AA collisions, considering 
the nuclear parton distributions discussed above, and estimate the ratio between these predictions and the unmodified pp production cross section. The so called nuclear modification ratios, defined in the Eqs. (\ref{fig:RpA}) and (\ref{fig:RAA}) below, could be measured at RHIC and LHC. We analyze the transverse momentum dependence of
the nuclear modification ratios. One have that if a proton-nucleus or nucleus-nucleus collision is nothing more than a superposition of nucleon-nucleon collisions, the ratios $R_{pA}$ and $R_{AA}$ should be unity. As the EKS and EPS sets are only evolved to
leading order (LO), the calculation of the cross sections is also done at LO
accuracy. To be consistent, we pick only the LO version of the other nPDFs we
are considering. Besides, since we are only calculating ratios between cross 
sections, common uncertainties on the normalization of the 
pA, AA and pp cross sections, due to higher order contributions, are expected to cancel
out in the ratios.

In Fig. \ref{fig:RpA} we present our estimates for the transverse momentum dependence of the ratio $R_{pA}$ defined by
\begin{eqnarray}
R_{pA}  \equiv { \frac{d\sigma (pA)}{dy d^2 p_T}|_{y=0} } / A {\frac{d\sigma
(pp)}{dy d^2 p_T}|_{y=0} } \,\,,
\end{eqnarray}
in pA collisions at RHIC ($\sqrt{s}=200\, GeV$) and LHC ($\sqrt{s}=8.8\, TeV$). 
The factorization scale $Q$ is assumed to be equal to the photon transverse momentum $p_T$.
The results show distinct behaviors of $R_{pA}$ for the distinct nPDF's and for the different colliders. The  difference  between the predictions  for $R_{pA}$ at RHIC and LHC comes from the kinematical $x_g$ range probed in the two cases. While for RHIC the values of $x_g$ are ever larger than $10^{-2}$, in the LHC case the minimum value is  $10^{-4}$, increasing with $p_T$. Consequently, the effects in the nuclear gluon distribution which contribute for the prompt photon production are different in the two colliders. It is verified in Fig. \ref{fig:RpA}, where we observe that at RHIC,  the EKS and EPS predictions implies $R_{pA}$ larger than one for $p_T\geq 5\, GeV$, being substantially larger for EPS, which is directly associated to the magnitude of the antishadowing present in $xg_A$. In contrast, for the DS and HKN predictions we have smaller values for the ratio due to the absence of this nuclear effect, with the ratio being larger for the DS prediction than the HKN one. At large $p_T$, which implies larger values of $x_g$,  the EKS and EPS predictions decreases due to the EMC effect. From these results we can conclude that the study of the ratio $R_{pA}$ at RHIC can be useful to determine the presence or not of the antishadowing and constrain its magnitude.
On the other hand, at LHC energies all parameterizations predict a ratio smaller than one, with  the suppression directly associated to the magnitude of the shadowing effect. For instance, we have that the ratio is substantially suppressed in the EPS (EKS) case, being smaller than $0.9$ (0.95) at $p_T \le 20$ GeV, while in the DS (HKN) case it is almost flat and equal to 0.95 (0.9) in the full $p_T$ range. Morever, differently from the DS and HKN predictions, the EKS and EPS one predict a  strong transverse momentum dependence. Consequently, the determination of the magnitude and $p_T$ dependence of the ratio at LHC  are useful to determine the properties of the shadowing in the nuclear gluon distribution.

\begin{figure}[t]
\includegraphics[scale=0.35]{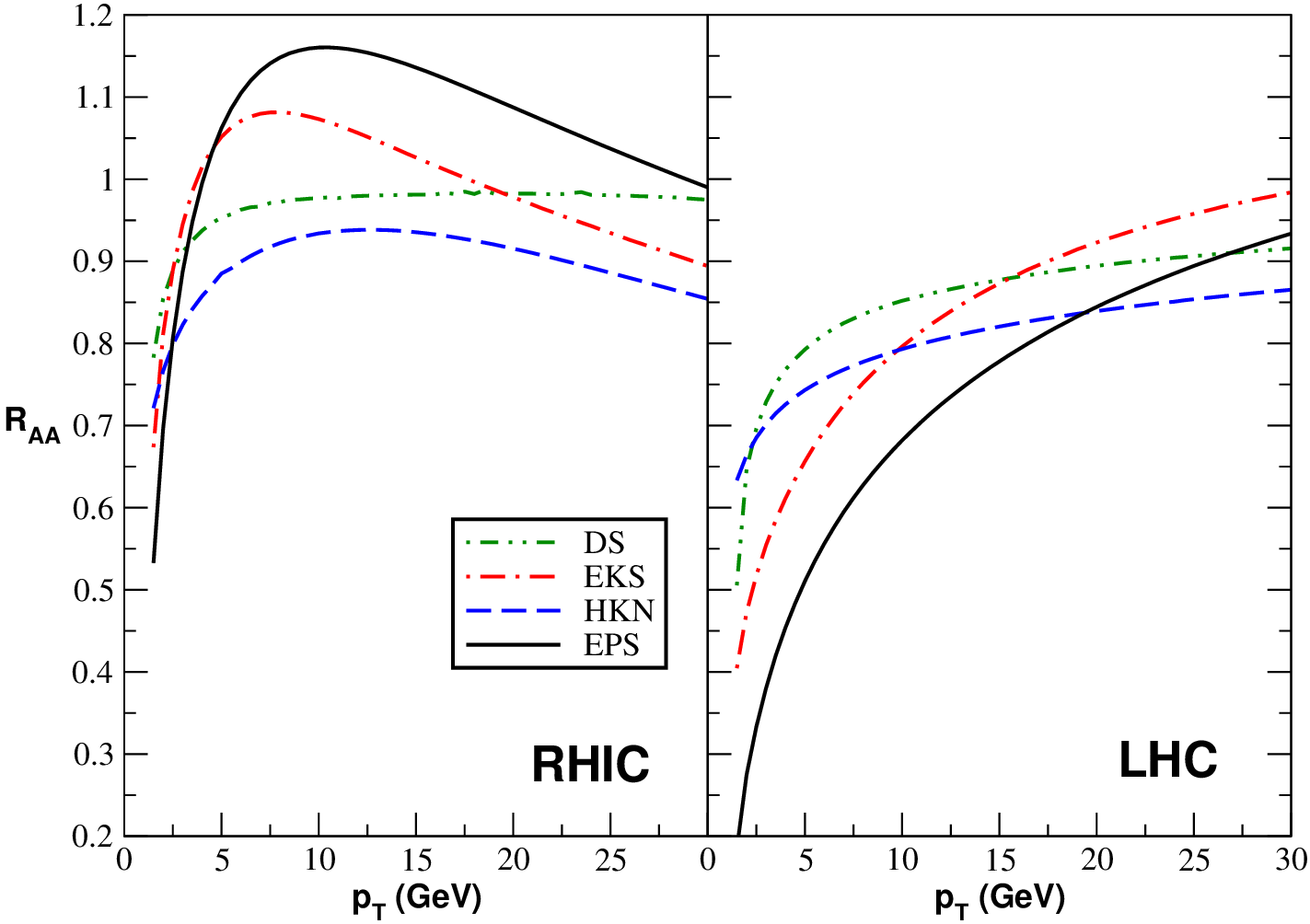}
\caption{Transverse momentum dependence of the ratio $R_{AA}$ 
in prompt photon production at RHIC ($\sqrt{s}=200\, GeV$) and LHC ($\sqrt{s}=5.5\, TeV$), for different choices of nuclear parton distributions:  DS \cite{sassot}, EKS \cite{EKS98}, HKN \cite{HKN} and EPS \cite{EPS08}. }
\label{fig:RAA}
\end{figure}

The production of photons can also be studied in the collision of heavy nuclei.  The main interest arises from the fact than, once they are produced, photons leave the hot and dense fireball virtually without further interactions (See e.g. Ref. \cite{hardprobes}). In particular, prompt photons constitute a physics background in the search for thermal photons, which are one of the QGP signatures. Therefore, a precise determination of the nuclear effects present in the initial state of the collision and the consequent suppression  of the cross section is  fundamental for the study of the photon production in ultrarelativistic heavy ion collisions at RHIC and LHC. In what follows  we present our estimates for the transverse
momentum dependence of the ratio $R_{AA}$ defined by
\begin{eqnarray}
R_{AA}  \equiv { \frac{d\sigma (AA)}{dy d^2 p_T}|_{y=0} } / A^2 {\frac{d\sigma
(pp)}{dy d^2 p_T}|_{y=0}} \,\,,
\end{eqnarray}
in AA collisions at RHIC  ($\sqrt{s}=200\, GeV$)  and LHC ($\sqrt{s}=5.5\, TeV$). In this case the nuclear effects are amplified by the presence of two nuclei in the initial state of the collision. Similarly to the $pA$ case, we can see in Fig. \ref{fig:RAA} that the ratio $R_{AA}$ is strongly modified by the shadowing and antishadowing effects. In particular, the  ratio is predicted by the EPS parameterization to be equal to 1.15 at $p_T = 10$ GeV and RHIC energies, while the DS parameterization predicts $R_{AA} \approx 1.0$ in the same $p_T$ range. Therefore, also in $AA$ collisions at RHIC, the study of prompt photon production can be used to constrain the magnitude of the antishadowing effect. On the other hand, at LHC energies we have that the ratio is strongly suppressed by the shadowing effect, with $R_g$ being predicted by the EPS parameterization to be approximately $0.5$ ($0.7$) at $p_T = 5.0$ (10) GeV . Our results indicate that the production of prompt photons with $p_T \le 20$ GeV can be used to determine the shadowing effect.

A comment is in order. In this letter we restrict ourselves to the descriptions which use the DGLAP evolution equations \cite{dglap} to describe the behavior of the nuclear parton
distributions. However, at small values of $x$, it is expected the presence of nonlinear corrections in the QCD evolution (For a recent review see \cite{hdqcd}). In this new regime, the nuclear gluon distribution should be determined by the solution of a nonlinear  evolution equation, as for example the  GLR-MQ evolution equation \cite{glr}. As shown in Ref. \cite{BMG:2007}, where the prompt photon production in $pp$ collisions was studied considering the solution of GLR-MQ evolution equation for the proton gluon distribution obtained in \cite{ehkqs}, new effects  are expected when  nonlinear PDFs are used in the calculations. A similar expectation is valid for the nuclear case.

In summary, in this letter we have investigated the prompt photon
production in pA and AA collisions at RHIC and LHC, considering the collinear factorization and some of the parameterizations for the nuclear parton distributions available in the literature.  Our results demonstrate that the nuclear ratios $R_{pA}$ and $R_{AA}$ are powerful observables to discriminate among the different parton distributions in the nuclear medium. In particular,  the predicted shadowing by the EPS parameterization is considerably larger than in the previous nuclear PDF's. This strong shadowing effect could be tested at the forthcoming LHC.

\begin{acknowledgments}
This work was  partially financed by the Brazilian funding agencies
FAPERGS and CNPq.
\end{acknowledgments}

\end{document}